\documentclass{aa}
\usepackage{psfig}

\def\ros{{\sl ROSAT }}
\def\ein{{\sl Einstein }}

\def\exo{{\sl EXOSAT }} 

\def\ari{{\sl Ariel V }}

\def\it{\sl}
\def\degs{\ifmmode ^{\circ}\else$^{\circ}$\fi}
\def\arcmin{\ifmmode ^{\prime}\else$^{\prime}$\fi}
\def\arcsec{\ifmmode ^{\prime\prime}\else$^{\prime\prime}$\fi}
\def\approxlt{\mathrel{\hbox{\rlap{\lower.55ex \hbox {$\sim$}}
        \kern-.3em \raise.4ex \hbox{$<$}}}}
\def\approxgt{\mathrel{\hbox{\rlap{\lower.55ex \hbox {$\sim$}}
        \kern-.3em \raise.4ex \hbox{$>$}}}}
\begin{document}
 
   \thesaurus{03         
              (11.01.2;  
               11.03.1;  
               11.09.1;  
               11.17.4;  
               13.25.2)  
}
   \title{The X-ray view of the quasar MR\,2251$-$178 and its host cluster: \\
          variability, absorption, and intracluster gas emission}  
   \author{Stefanie Komossa} 

   \offprints{St. Komossa, skomossa@mpe.mpg.de  }
 
  \institute{Max-Planck-Institut f\"ur extraterrestrische Physik, Postfach 1312,   
             D-85741 Garching, Germany\\
       }

   \date{Received  25 October 2000; accepted December 2000}

   \maketitle\markboth{St. Komossa, The X-ray view of MR\,2251$-$178 and its environment}{}  

   \begin{abstract}
MR\,2251-178 was the first quasar initially discovered in X-rays,
and the first one found to host a warm absorber.
The quasar turned out to be an outstanding object in many respects.
It has a high ratio of X-ray/optical luminosity, is surrounded
by the largest quasar emission-line nebula known, and
is located in the outskirts of a cluster of galaxies. 
Here, we present results from an
analysis of the X-ray spectral, temporal, and spatial
properties of this source and its environment
based on deep {\sl ROSAT} observations.

Remarkably, we do not detect any excess X-ray {\em cold} absorption 
expected to originate from the giant gas nebula surrounding MR2251-178. 
This excludes the presence of a huge HI envelope
around the quasar. 
The X-ray spectrum of MR2251-178 is best fit by a warm-absorbed
powerlaw  with
an ionization parameter
$\log U = 0.5$ and a column density
$\log N_{\rm w} = 22.6$ which, however, cannot be the same material
as the giant optical emission line nebula.
The mean (0.1--2.4)keV X-ray luminosity amounts to $10^{45}$ erg/s.  

A spatial analysis shows that the bulk of the X-ray emission from
the quasar is consistent with a point source, as expected in view of the
powerlaw-shaped X-ray spectrum and the rapid X-ray variability we detect.
In addition, extended emission appears at weak emission levels,
including a bridge between the quasar and the cD galaxy of the cluster. 

The X-ray emission from the intra-cluster medium is
weak or absent. We derive an upper limit on the X-ray luminosity
of $L_{\rm x} \le 1.6\,10^{42}$ erg/s,  weaker than other clusters
of comparable richness.   
None of the other member galaxies of the cluster to which
MR\,2251-178 belongs, are detected in X-rays. However,
east of the quasar there is a significant excess of X-ray sources, 
several of them without optical counterparts on the UK Schmidt plates.
      

   \end{abstract}
 
\section{Introduction}

The quasar MR\,2251-178, at a redshift of $z$=0.064, 
is located in the outskirts of an irregular cluster of galaxies
(Phillips et al. 1980). 
It was the first quasar initially identified from X-ray observations  
(Cooke et al. 1978, Ricker et al. 1978). 
MR\,2251-178 shows remarkable multi-wavelength properties.  

In the optical spectral region, much attention has focussed on the 
very extended [OIII] emission line region of MR\,2251-178 found by Bergeron et al. (1983),
and the source of its ionization and its origin (e.g., di Serego Alighieri et al. 1984,
Hansen et al. 1984, Macchetto et al. 1990).
The quasar is a weak radio emitter of Fanaroff-Riley type I
(Kembhavi et al. 1986, Macchetto et al. 1990).  

MR\,2251-178 was detected as a bright X-ray source 
in the course of the 
\ari all-sky survey (Cooke et al. 1978) and identified with the quasar 
by Ricker et al. (1978) from observations with the SAS-3 X-ray observatory.
The X-ray spectrum of MR\,2251-178 was first studied by Halpern (1984) who 
deduced the presence of a warm absorber based on the observed 
variability pattern of the \ein data. 
Based on {\sl EXOSAT} and {\sl Ginga} observations
Pan et al. (1990) and Mineo \& Stewart (1993), again,
interpreted and modeled the data in terms of the presence of a warm absorber.
In particular, Mineo \& Stewart found the ionization state of the absorber
to follow changes in the intrinsic continuum luminosity.
An independent analysis of the \exo data was performed by
Walter \& Courvoisier (1992), who concluded that the previously reported
variability could be completely traced back to variability
in the powerlaw index (and {\em constant cold} absorption)
rendering the presence of a warm absorber unnecessary.
Reynolds (1997; see also Reeves et al. 1997),
in the course of a large sample
study of {\sl ASCA} observations of AGN, reported the detection
of OVII and OVIII absorption edges in the X-ray spectrum of MR\,2251-178.

Warm absorbers are an important diagnostic of the
physical conditions
within the central regions of active galaxies.
They have been observed in $\sim$50\% of the well-studied Seyfert galaxies
(see Komossa 1999 for a review). 
Their presence was recently confirmed
with {\sl Chandra} observations of NGC\,5548 which reveal a rich absorption
line spectrum (Kaastra et al. 2000). 
The study of the ionized material provides a wealth of information
about the nature of the warm absorber itself, its relation
to other components of the active nucleus, and the intrinsic
AGN X-ray spectral shape, and leads to a
 better  understanding of AGN in general.
Thorough modeling of the few {\em quasars} that show warm absorbers,
like MR\,2251-178, is also important for investigating
the relation between warm absorbers in Seyferts and quasars
and the cause for their different abundances.

Here, we present results from an 
analysis of a deep archival {\sl ROSAT} PSPC
(Tr\"umper 1983) observation of
MR\,2251-178. In addition, {\sl ROSAT} all-sky survey observations
of this source 
are analyzed in order to study the long-term
variability.
Besides an investigation of the properties of MR\,2251-178 itself, the following issues
are of importance: 
the potential contribution of nearby 
sources to the X-ray spectrum of MR\,2251-178, 
the search for X-ray emission from other member galaxies
of the cluster to which MR\,2251-178 belongs, and the search
for emission from the intra-cluster medium. 
The high spatial resolution of the {\sl ROSAT} observation
allows for the first time to separate these components.  

The results presented here were obtained in the course of
a systematic study of bright \ros AGNs that show X-ray spectral
complexity, including NGC\,4051 (Komossa \& Fink 1997a),
NGC\,3227 (Komossa \& Fink 1997b), 
IRAS\,17020+4544 (Komossa \& Bade 1998),
IRAS\,13349+2438 (Komossa \& Breitschwerdt 2000, Siebert et al. 1999), and 
the spectrally highly variable Narrow-line Seyfert\,1 galaxy RXJ\,0134-4258
(Komossa \& Meerschweinchen 2000). 

The paper is organized as follows: 
In Sect. 2 we present the observations. 
The analysis of the data with respect to their spatial, temporal,
and spectral properties 
is given in  
Sects. 3, 4, and 5, respectively. 
Sect. 6 presents the discussion,  
and a summary and the conclusions are provided in Section 7. 
A distance of 410 Mpc was adopted for MR\,2251-178.
If not stated otherwise, cgs units are used throughout the paper.

\section{Data reduction}

\subsection{Pointed observation} 
The on-axis observation of MR\,2251-178 
was performed with the \ros PSPC on Nov. 15--16, 1993.  
The exposure time was 18.3 ksec.   
In total, 36~X-ray sources were detected with a 
likelihood $\ge$ 15 within the PSPC field of view.
The target-source photons were extracted
within a circle centered around the X-ray position of MR\,2251-178. 
The background was determined from the inner 19\arcmin~ of the field of view, 
after the removal of all detected sources.   
Vignetting
and dead-time corrections were applied to the data 
using the EXSAS software package (Zimmermann et al. 1994).
The mean source count rate is 3.10$\pm{0.05}$ cts/s.
For the spectral analysis source photons in
the amplitude channels 11-240 were binned 
according to a signal/noise ratio of 35$\sigma$.

\subsection{ROSAT all-sky survey data}

The sky field around MR\,2251-178 was observed during the \ros all-sky survey (RASS)
in November 1990. 
We determined the background from a source-free region along
the scanning direction of the telescope
and corrected the data for vignetting. 
Emission from MR\,2251-178 is clearly detected. We derive a count rate of 
1.0 cts/s, about a factor of 3 weaker than during the pointed observation.  

Results presented below refer to the pointed observation if not
noted otherwise. The RASS data are mainly used to determine
the long-term variability.   

\begin{figure} 
 \psfig{file=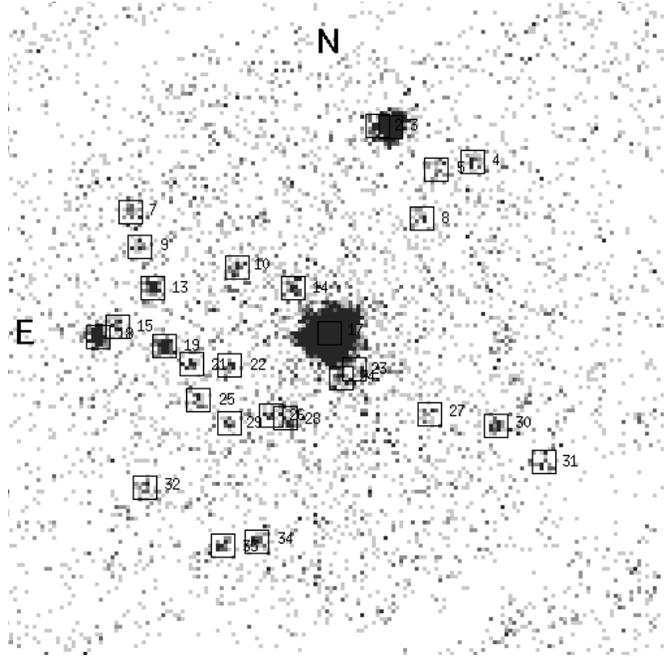,width=8.7cm,clip=}
 \caption[x] {X-ray sources detected in a 40\arcmin~$\times$ 40\arcmin~field
around MR\,2251-178.   }
\end{figure}
\begin{figure} 
 \psfig{file=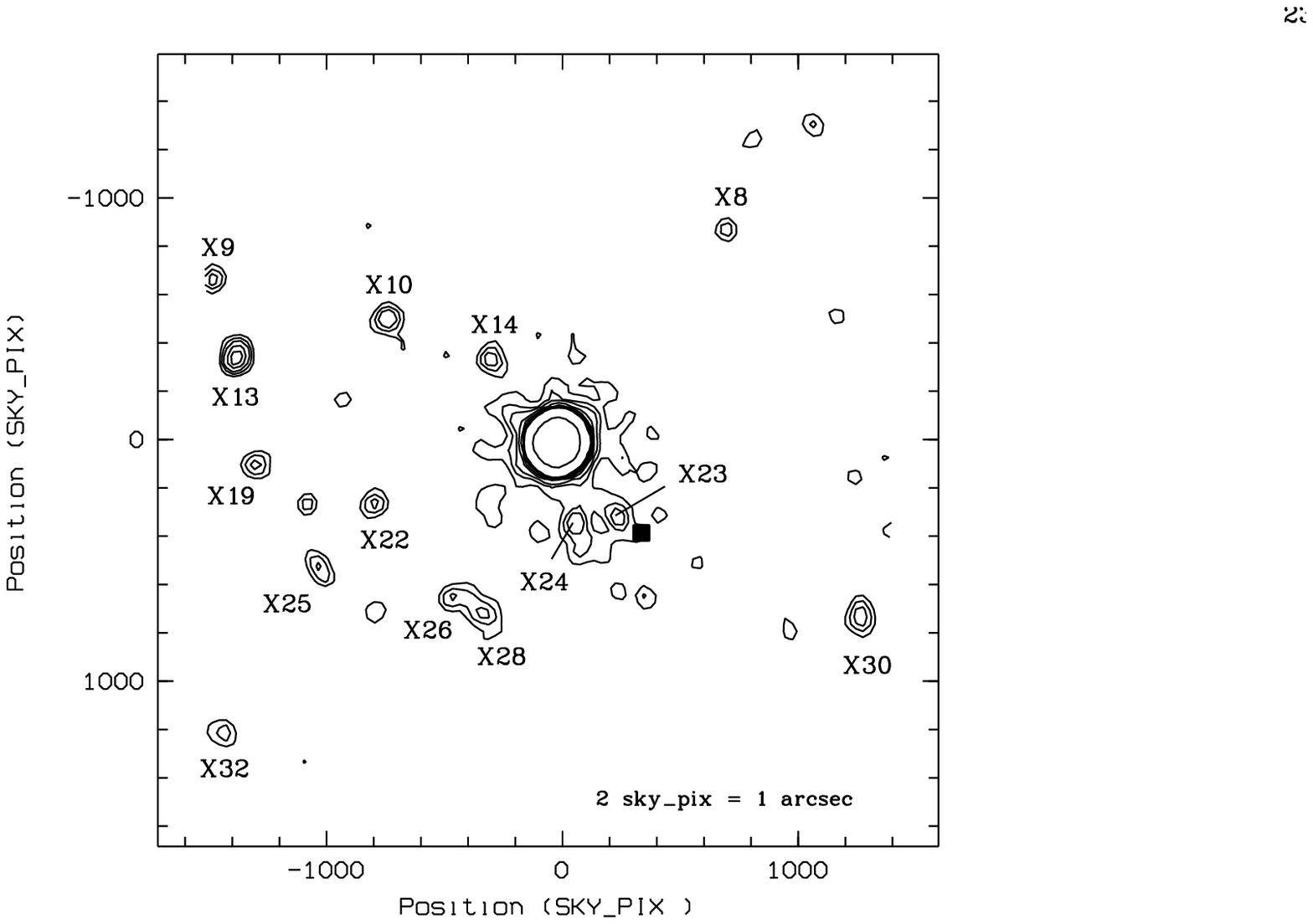,width=8.7cm,clip=}
 \caption[] {Contour plot 
 of the field around MR2251-178 in the energy range 0.5--1.5 keV. 1 sky-pix
corresponds to a scale of 0.5\arcsec. The brightest X-ray sources are labeled.
The contours are plotted at 
1,1.5,2-8 and 100$\sigma$ above the background. The filled square marks
the location of the cD galaxy of the cluster to which MR2251-178 belongs.}
\end{figure}

\section{Spatial analysis}

36~X-ray sources were detected with
a likelihood $\ge$ 15 within the PSPC field of view (Fig. 1).
The positions of those in the vicinity of MR\,2251-178
are shown in Fig. 3, overlaid
on an optical image from the UK Schmidt plates.
The nearby sources are weak (see Tab. 1 for
count rates); none is expected to have significantly
confused previous X-ray observations of MR\,2251-178.
Count rates of all sources detected in the inner PSPC field
of view (shown in Fig. 1) are listed in Tab. 1. They
are of interest for later variability studies if MR2251-178
is re-observed by {\sl Chandra} or {\sl XMM}. 

 \begin{table*}     
     \caption{Count rates of X-ray sources detected around MR2251-178,
in the inner PSPC field of view. Individual sources are labeled with 'X'
followed by their number as in Figure 1. 
Powerlaw spectral fits to the brightest source, X3, and the brightest
of the eastern-excess sources (see Sects. 5.2, 6.4), X18, give
$\Gamma_{\rm X3} = -2.8\pm{0.1}$ and $\Gamma_{\rm X18} = -3.1\pm{0.2}$,
respectively, for $N_{\rm H}$ fixed to $N_{\rm Gal}$.  
}
  \label{crates}
  \begin{tabular}{ccccccccccccccc}
      \hline
      \noalign{\smallskip}
  source &                     X3  & X4 & X5 & X7 & X8 & X9 & X10 & X13 & X14 & X15 & MR\,2251 & X18 & X19 &
                  X21 \\
  count rate [10$^{-3}$cts/s] & 57 & 3.5& 3.4& 2.9& 1.9& 2.7& 3.9 & 7.1 & 3.7 & 3.0 & 3100       & 13.9& 7.6 &
                  2.7 \\
      \noalign{\smallskip}
      \noalign{\smallskip}
      \noalign{\smallskip}
  source &                     X22 & X23 & X24 & X25 & X26 & X27 & X28 & X29 & X30 & X31 & X32 & X34 & X35 & \\                    
  count rate [10$^{-3}$cts/s] & 2.3 & 9.4 & 3.5 & 2.3 & 2.2 & 1.7 & 5.5 & 2.1 & 4.3 & 2.5 & 3.1 & 3.9 & 3.4 & \\  
      \noalign{\smallskip}
      \hline
  \end{tabular}

   \end{table*}

We note in passing that two
of the three X-ray sources shown in Fig. 3 are aligned
with MR\,2251-178 and with the bright elliptical galaxy. 
While the present case could be just
coincidence, we refer the reader to
Arp (1997, and references therein) for a thorough 
analysis of alignments of X-ray sources around nearby galaxies.

None of the other member galaxies (e.g., Fig.\,2 of Phillips 1980)
of the cluster to which
MR\,2251-178 belongs, are detected in X-rays. However,
there is an excess of X-ray sources east of the quasar (Fig. 1).
Several
of these sources do not have optical counterparts on the UK Schmidt plates.
A correlation of the X-ray positions of all 36 detected sources
with the SIMBAD database results in only one identification,
MR\,2251-178.

A spatial analysis of the X-ray emission from MR2251-178 shows
that the bulk of the X-ray emission is consistent with emission from
a point source. We find a systematic broadening of the radial source profile
of order 1-3$\arcsec$ in comparison with the core of the theoretical
point spread function of the instrument. Such deviations have been
observed previously in other sources, and can be traced back to 
instrumental effects (e.g., Morse 1994). 
Figure 2 provides a contour plot of the X-ray emission in the
field around MR2251-178. At weak emission levels, extended emission
appears. In particular, there is a bridge-like feature
extending from the quasar in the direction of the cD galaxy
of the cluster (the feature does {\em not} point in the direction of
the wobble motion of the satellite).  

Finally, we searched for extended X-ray emission from the intra-cluster
medium. Phillips et al. (1980) locate the center of the cluster of galaxies
close to the bright elliptical galaxy which can be seen in Fig. 3.  
We first removed all sources detected with a likelihood $l \ge 10$
in the PSPC field of view. 
In the next step, eight source-free background regions symmetrically
distributed around the target region,
but in the outer parts of the field of view, were selected and the mean
background count rate was determined. We then derived the background-corrected number
of photons in a circular region of 150\arcsec~radius, 
with its position close to the optical center of the cluster of galaxies.   
A number of 89 excess photons is detected. These could either originate
from unresolved point sources, or represent emission from the intracluster
gas. We conservatively assume that all excess photons originate from
the intracluster medium (ICM) in order to derive an upper limit on its X-ray
emission. To convert the countrate to X-ray luminosity, we used
a Raymond-Smith emission model and assumed
an ICM temperature of (i) $kT$ = 5 keV (typical for clusters of galaxies)
and (ii) 1.5 keV (more typical for groups of galaxies).  
This yields X-ray luminosities of $L_{\rm x_i} \le 1.6\,10^{42}$ erg/s and 
$L_{\rm x_{ii}} \le 1.4\,10^{42}$ erg/s, respectively.

\begin{figure}[th]
\psfig{file=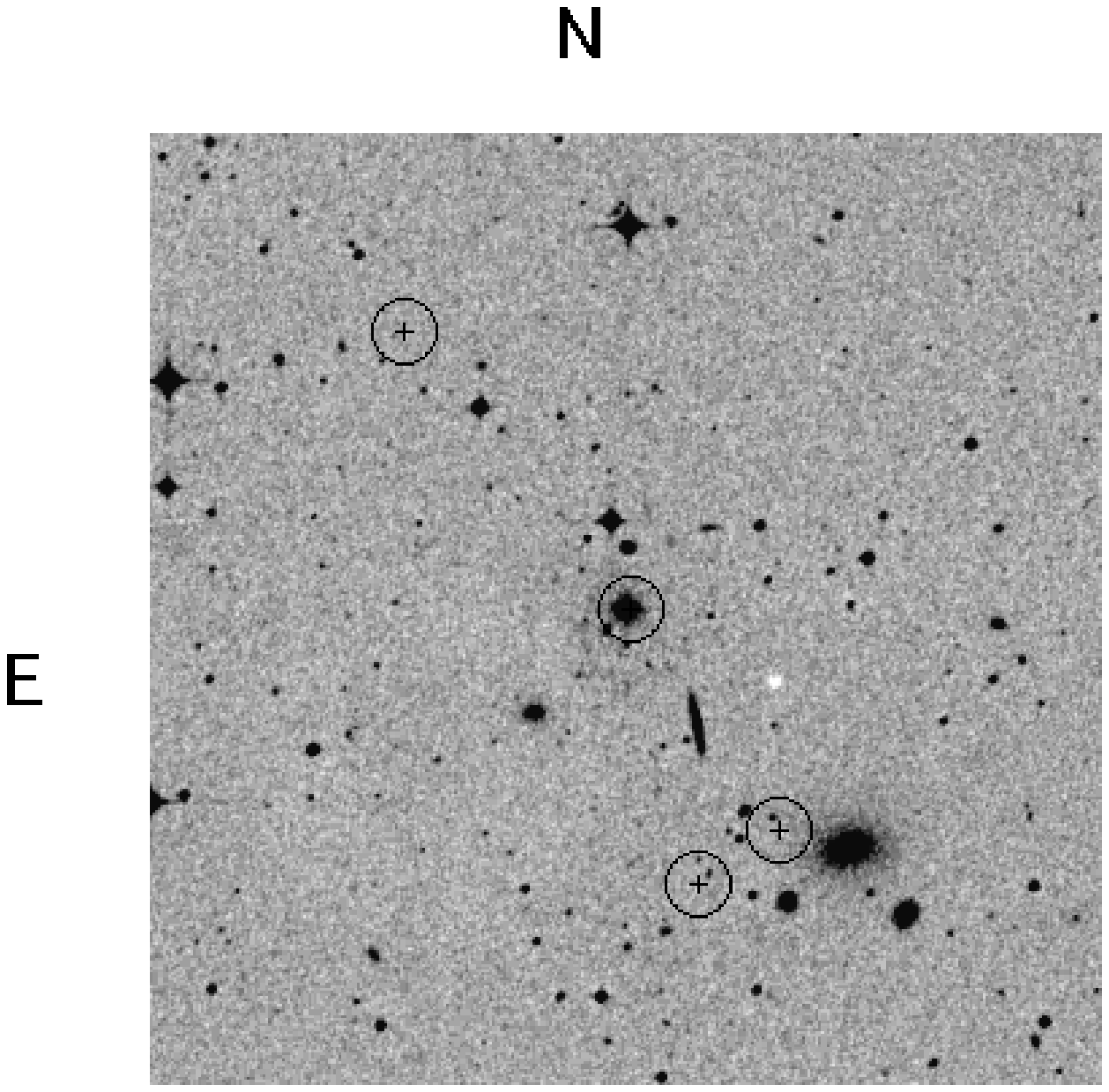,width=8.7cm,clip=}
 \caption[mr_ima]{X-ray sources detected
in a 10\arcmin~$\times$ 10\arcmin~field
around MR\,2251-178 superimposed on an optical image.
The circles drawn around
the X-ray source positions are of 20\arcsec~radius.
   }
\end{figure}

\section{Temporal analysis}

The mean source count rate during the pointed observation
was 3.1 cts/s, a factor
of 3 higher than during the {\sl ROSAT} all-sky
survey (RASS) observation performed 3 years earlier.
The X-ray lightcurves are displayed in Fig. 4.
We find evidence for a remarkable flaring
event with a rise time of 10\,000 sec and an amplitude
of a factor 2 in count rate  which ocurred during
the RASS (Fig. 4, upper panel).
It is interesting to mention here the luminous quasar PDS 456,
where a similar flaring event (factor 2 rise in countrate within 17\,ksec)
was seen 
(Reeves et al. 2000).  

In order to study variability during the pointed observation,
in a first step a time binning of 400\,s was adopted
to account for the wobble motion of the satellite
(Fig. 4, lower panel).
In a second step, we chose time bins of 20s. On average,
this still gives $\sim$60 source counts per bin. The
wobble-induced source variability is expected to be less
than $\sim$30\%. We find repeated short-term variability by
up to a factor 2 in count rate 
(not visible in the lower panel of Fig. 4 where time bins were
400s each), 
unexpected for such a luminous
quasar.

\begin{figure} 
\psfig{file=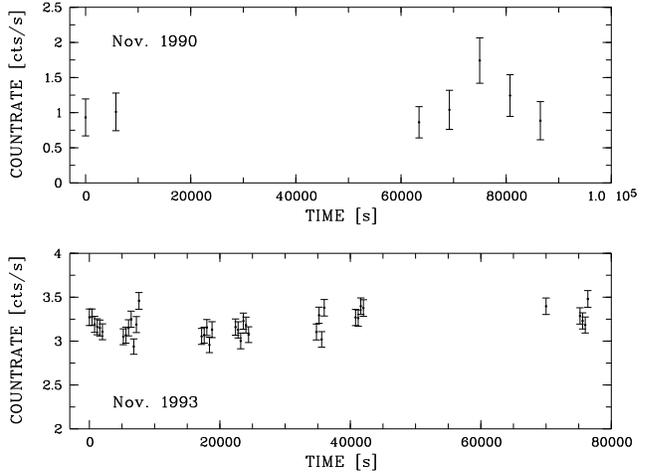,width=9cm}
\caption[mr2251_light]{X-ray lightcurve of MR\,2251-178
during the {\sl ROSAT} all-sky survey (upper panel)
and during the later pointed PSPC observation (lower panel;
time bins of 400 sec).
 }
\end{figure}

\section{Spectral analysis} 

\subsection{MR\,2251-178}

\subsubsection{Standard spectral models}

First, we fit a single powerlaw to the X-ray spectrum.
The column density of cold absorbing material
was fixed to the Galactic value towards MR\,2251-178,
$N_{\rm Gal} = 2.77 \times 10^{20}$ cm$^{-2}$ (Lockman \& Savage 1995),
since it is underpredicted otherwise (Fig. 5).
This gives $\Gamma_{\rm x}=-2.3$ and clearly is a very poor description of the
data ($\chi^2_{\rm red}=5.4$; Fig. 6).
%
\begin{figure} 
 \psfig{file=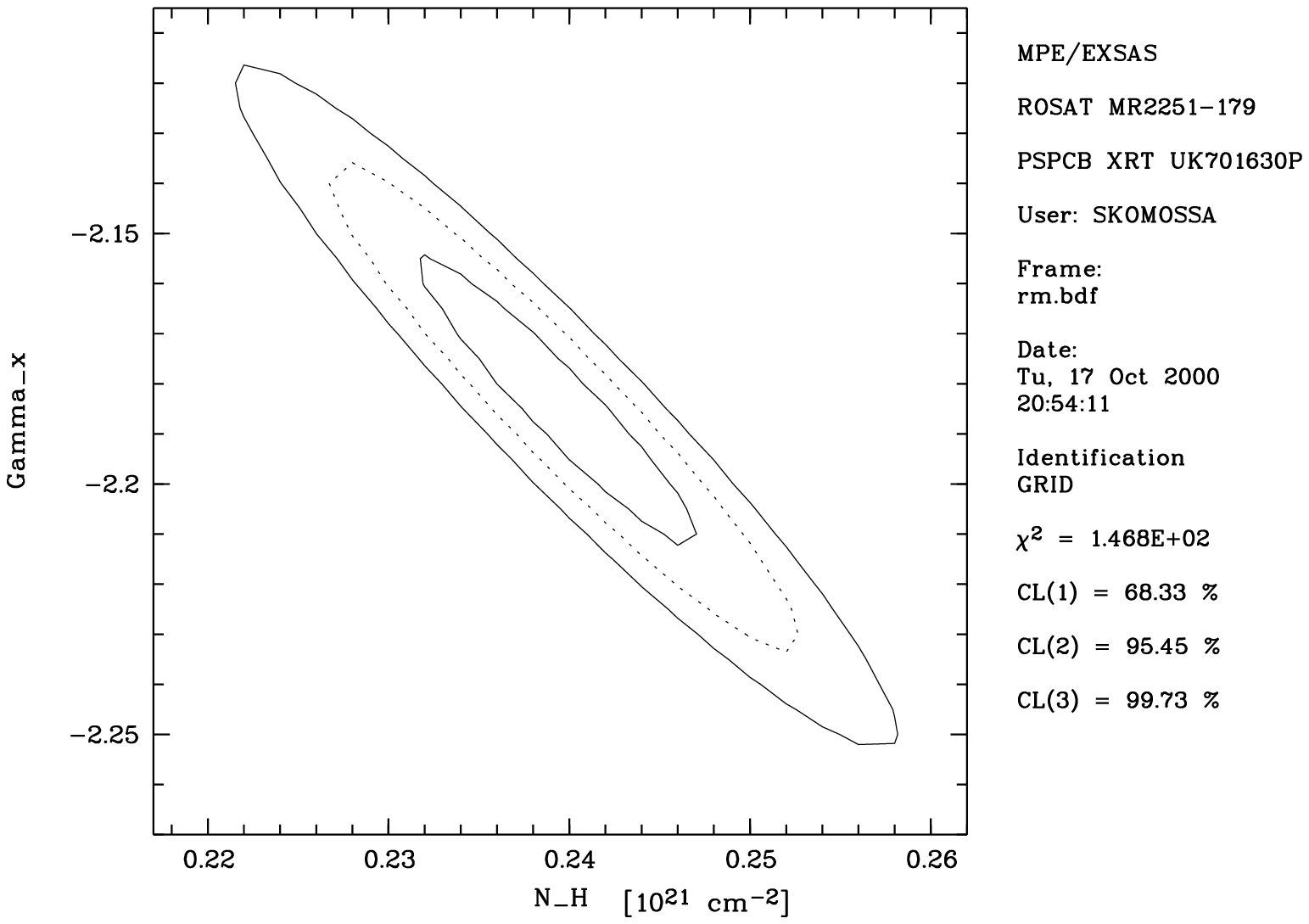,width=8.4cm,clip=}
 \caption[x] {Error ellipses in $\Gamma_{\rm x}$, $N_{\rm H}$ for a powerlaw fit to MR2251-178.
The two dimensional contours are shown at confidence levels
of 68.3, 95.5 and 99.7\%.} 
\end{figure}
A Raymond-Smith emission model does not fit the spectrum
at all, even if temperature, metal abundances
and the amount of cold absorption are all treated as free
parameters.
We then tried two-component spectral fits involving a powerlaw
plus soft excess parametrized by different models. The
quality of the fit remains unacceptable, though (Table 2).{\footnote{In 
particular, we checked whether the combination of
excess X-ray absorption and a very soft black body component
 -- compensating each other to mimic an unabsorbed spectrum -- 
is a possibility (see below for the motivation to
search carefully for excess X-ray cold absorption). 
If both, the black body temperature and the absorption
are treated as free parameters, we find $N_{\rm H} = 0.24\pm{0.05}$ 10$^{21}$ cm$^{-2}$;
less than the Galactic value, but consistent within the errors. 
In a second step, we enforced excess absorption but were unable
to obtain any successful spectral fit. Finally we note that 
we repeated the fits of all models listed in Table 2
with $N_{\rm H}$ as free parameter; it is never found to exceed
the Galactic value. This still holds if we ignore in the fitting procedure
the first three 
spectral bins  which always show a positive deviation from the
best fit model.}} 
We have also experimented with a powerlaw plus two gaussian
emission lines to account for the positive residuals
around $\sim$ 0.55 and 1.8 keV of the single powerlaw fit
(Fig. 6, middle panel). However, to match the rather similar
widths of the residuals despite the different instrumental
broadening at these two energies, two very different lines are
required; the first one at 0.55 keV extremely narrow, the
second one at 1.8 keV very broad with a width of $\sim$500 eV.
We therefore do not discuss this model further.

   \begin{table*}     
     \caption{X-ray spectral fits to MR\,2251-178 (pl = powerlaw, bb = black body,
                  wa = warm absorber, RS = Raymond-Smith emission model).  
                  The errors are quoted at the 90\% confidence level. The first
                  row of results refers to the {\sl ROSAT} survey observation (RASS),
                 all others refer to the pointed observation.  }
     \label{fitres}
      \begin{tabular}{lllclcc}
      \hline
      \noalign{\smallskip}
        No. & model & $N_{\rm H}$ & wa parameters  
                            & ~$\Gamma_{\rm x}$ & $kT$
                            & $\chi^2_{\rm red}$ \\
       \noalign{\smallskip}
      \noalign{\smallskip}
         &   & [10$^{21}$ cm$^{-2}$] & & & 
                  keV &  \\
       \noalign{\smallskip}
      \hline
      \hline
      \noalign{\smallskip}
 (1) & pl, RASS & 0.277$^{(1)}$ & - & --1.9$\pm{0.3}$ & - & 1.4 \\
      \noalign{\smallskip}
      \hline
      \noalign{\smallskip}
 (2) & pl & 0.277$^{(1)}$ & - & --2.32$\pm{0.01}$ & - & 5.4 \\
      \noalign{\smallskip}
      \hline
      \noalign{\smallskip}
 (3) & pl & 0.24$\pm{0.01}$ & - & --2.18$\pm{0.04}$ & - & 4.2 \\
      \noalign{\smallskip}
      \hline
      \noalign{\smallskip}
 (4) & RS$^{(4)}$ & 0.124$\pm{0.002}$   & - & - & 1.4$\pm{0.1}$ & 13.4 \\
      \noalign{\smallskip}
      \hline
      \noalign{\smallskip}
 (5) & pl + bb &  0.277$^{(1)}$ & - & --2.20$\pm{0.20}$ & 0.4$\pm{0.3}$ & 4.5 \\
      \noalign{\smallskip}
      \hline
      \noalign{\smallskip}
 (6a) & pl + 2\,edges & 0.277$^{(1)}$ & $\tau_{\rm OVII}$=0.26$\pm{0.12}$, $\tau_{\rm OVIII}$=0.20$\pm{0.12}$
                           & --2.23$\pm{0.02}$ & - & 1.6 \\
 (6b) & pl + 2\,edges$^{(3)}$  & 0.277$^{(1)}$ & $\tau_{\rm OVII}$=0.22$\pm{0.11}$, $\tau_{\rm OVIII}$=0.24$\pm{0.12}$
                           & --2.21$\pm{0.02}$ & - & 0.9 \\ 
      \noalign{\smallskip}
      \hline
      \noalign{\smallskip}
 (7) & wa$^{(3)}$ & 0.277$^{(1)}$ & $\log U = 0.5^{+0.3}_{-0.2}$, $\log N_{\rm w} = 22.6^{+0.2}_{-0.1}$  
                & -1.9$^{(2)}$ & - & 1.0 \\
      \noalign{\smallskip}
      \hline
      \noalign{\smallskip}
  \end{tabular}

\noindent{\small $^{(1)}$ fixed to the Galactic value, $^{(2)}$ fixed, 
            $^{(3)}$ first three bins excluded from fit, $^{(4)}$ abundances fixed to 0.01\,solar; fit gets worse
            for higher abundances 
}
   \end{table*}

\begin{figure}[t]
 \psfig{file=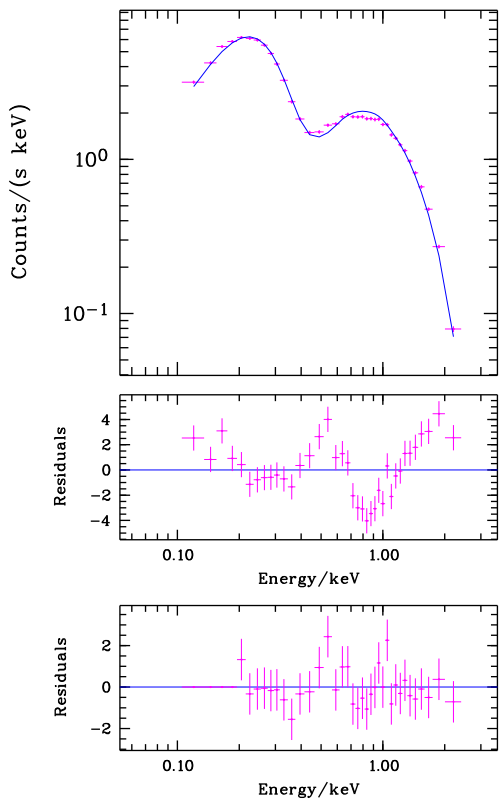,width=7.0cm}
\caption[mr_x]{The upper panel shows the observed X-ray spectrum of MR\,2251-178 (binned
to S/N=35; crosses)
and the best-fit powerlaw model (solid line).
The second panel displays
the fit residuals for this model, whereas the lowest panel gives
the residuals from a warm absorber fit 
(note the different scale of the ordinate).
The lowest energy bins were ignored in this fit (see text for details).  }
\end{figure}

\subsubsection{Warm absorber models}

The presence of a warm absorber markedly improves
the fit. Performing a two-edges fit with edge energies fixed
at the theoretical values of OVII and OVIII we obtain
$\tau_{\rm OVII}=0.26$, $\tau_{\rm OVIII}=0.20$,
and $\Gamma_{\rm x} = -2.20$.
The not yet totally satisfactory quality of the fit,
$\chi^2_{\rm red}=1.6$, can be traced back to a deviation of
the low energy part of the spectrum (between 0.1--0.18 keV)
which might be caused by residual calibration uncertainties,
or a new spectral component of which we just see the tail.
If this part of the spectrum is excluded from the spectral
fitting, we obtain $\chi^2_{\rm red}=0.9$, $\tau_{\rm OVII}=0.22$,
$\tau_{\rm OVIII}=0.24$, and $\Gamma_{\rm x} = -2.21$.

Next, a warm absorber model based on photoionization
calculations (Komossa \& Fink 1997a)
with the code {\em Cloudy} (Ferland 1993) was
applied. 
The ionized material was assumed to be photoionized by the continuum emission of the
nucleus, to be one-component and of constant density.
Solar abundances (Grevesse \& Anders 1989) were adopted.
Outside the soft X-ray band, the spectral energy distribution incident on the warm material
was chosen to be a mean Seyfert continuum taken from
Komossa \& Schulz (1997){\footnote{The SED consists of
an UV-EUV powerlaw of energy index $\alpha_{\rm uv-x}$=--1.4
extending up to 0.1 keV, a mean optical to radio continuum after
Padovani \& Rafanelli (1988), a break at 10$\mu$m and an index
$\alpha$ = --2.5 $\lambda$-longwards. 
For comments on the
usually weak influence of non - X-ray spectral parts on the warm absorption
structure see Komossa \& Fink (1997a).
The soft X-ray continuum was directly determined from spectral
fits ($\Gamma_{\rm x}$ = --1.9)
and extrapolated to higher energies, with a break at 100 keV.}}.
The two properties characterizing the warm absorber 
that can be directly extracted from X-ray spectral fitting 
are the hydrogen column density $N_{\rm w}$ of the ionized material and the     
ionization parameter $U$. The latter is defined as  
$U=Q/(4\pi{r}^{2}n_{\rm H}c)$, 
where $Q$ is the
number rate of incident photons above the Lyman limit, $r$ is the distance between
nucleus and warm absorber, $n_{\rm H}$ is the hydrogen number density 
and $c$ the speed of light.  

A fit of the X-ray spectrum of MR\,2251-178 gives
an ionization parameter of the warm absorber
$\log U = 0.5$ and a column density
$\log N_{\rm w} = 22.6$.
We then split the total data set in three subsets according
to flux-state of the source, and fit these data separately.
The best-fit para\-meters remain the same within the error bars;
no changes in the warm absorber are detected.
This also holds for the RASS observation performed 3 yrs earlier;
the best-fit column density is a factor 3 lower, but consistent with being
constant within the errors.  

The best-fit warm absorber model still leaves some residuals
at the low energy-part of the spectrum, suggesting the presence
of a second, lower ionized absorber which could possibly be dusty
(see Komossa 1999 for a review on dusty warm absorbers).
These properties make MR\,2251-178 an excellent candidate for
observations with the LETG spectrometer aboard {\em Chandra}.

Integration over the best-fit X-ray spectrum (we used model 6a of Table 2)
after correction for cold and warm absorption
gives a mean rest-frame luminosity of $L_{\rm (0.1-2.4) keV} = 1.4\,10^{45}$ erg/s. 
MR\,2251-178 is thus one of the most luminous soft X-ray emitters in the local ($z<$0.1)
universe. 

\subsection{X18}

In order to get first clues on the identity of the other X-ray sources,
we have analyzed the source X18
which is the brightest of the eastern-excess sources.
$N_{\rm H}$ was fixed to the same value used
for MR2251-178. We repeated all fits for $z=0.064$ and zero redshift. 

A powerlaw spectral fit to 
X18 gives $\Gamma_{\rm x} = -3.1\pm{0.2}$ ($\chi^2_{\rm red}=1.2$),
a Raymond-Smith fit yields $\chi^2_{\rm red}=9$, a single black
body model is similarly unsuccessful ($\chi^2_{\rm red}=4.7$). 
If X18 is located at the distance of MR2251-178 its 
luminosity is $L_{\rm x} = 9\,10^{42}$ erg/s, based on the powerlaw fit.

\section{Discussion}

\subsection{Warm and cold absorption, and relation to the giant gas nebula surrounding MR2251-178}

MR2251-178 is one of the few quasars that host warm absorbers.
Our simple one-component warm absorber
fit gives a high column density of the ionized material, similar to that 
seen in NGC\,4051 (e.g., Komossa \& Fink 1997a).
We have checked that none of the observed optical-UV emission lines
of MR2251-178 are overpredicted by our best-fit warm absorber model. 
For instance, we derive a warm-absorber intrinsic intensity ratio
in the emission lines [FeXIV]$\lambda$5303/H$\beta$ of 0.02.  

It is interesting to note that the optical emission-line 
gas which surrounds MR2251-178 
is of huge hydrogen column density.
Essentially, two components have been distinguished: a circumnuclear
component with an extent of $\sim$20 kpc, and a giant envelope
which extends up to $\sim$200 kpc. 
Bergeron et al. (1983) provided crude estimates of the mean {\em ionized} gas densities 
of $n \approx 0.3$ cm$^{-3}$  and $n \approx 0.01$ cm$^{-3}$ for
the circumnuclear component and the envelope, respectively, with uncertainties
larger than a factor 10.
This translates into column densities of 9\,10$^{21}$ cm$^{-2}$ and 3\,10$^{21}$ cm$^{-2}$.    
If this gas were neutral, the quasar would be heavily absorbed
at soft X-ray energies. 
We note in passing that this material is not expected to act as 
X-ray warm absorber, since its degree of ionization is too low. 
In our best-fit warm absorber model, the oxygen ion species 
of lowest ionization with significant abundance is O$^{5+}$.  

Several scenarios have been proposed to explain the 
origin of the ionized gas envelope around MR2251-178,
including a merger remnant as result of a merger event between 
MR2251-178 and a small, gas-rich spiral galaxy;
tidally stripped gas from the neighbour galaxy; 
the ionized part of a large HI envelope, possibly
left over from the formation epoch of the quasar;
or material expelled from the QSO  (e.g., 
Bergeron et al. 1983, Norgaard-Nielsen et al. 1986,
Macchetto et al. 1990).
Based on the extent, symmetry, and rotation pattern
of the envelope, Shopbell et al. (1999) conclude that the interaction
model is unlikely and that the envelope most probably did not
originate within the host galaxy of the quasar.  
They favor a model in which the extended
HII envelope resides within a large complex of HI gas 
centered about the quasar.
Is has repeatedly been suggested to search for such an HI halo 
using the 21cm line (e.g., Bergeron et al. 1983, Macchetto et al. 1990,
Shopbell et al. 1999). 

Alternative to radio observations, soft X-rays are an excellent 
probe of excess absorption
from cold material along the line of sight. 
Such absorption is expected, firstly, from the HII gas component itself 
(more likely from the envelope than the circumnuclear component)
if more than $\sim$1\% of H is neutral,  
and, secondly, from the HI envelope.  
We have carefully searched for such absorption, 
and do not find any evidence for it.{\footnote{If we 
increase the amount of absorbing material by $N_{\rm H} = 5\,10^{19}$ cm$^{-2}$
and re-fit the X-ray spectrum, the fit is drastically worse, with a change
$\Delta \chi^2$ = 187. We therefore consider $N_{\rm H} = 5\,10^{19}$ cm$^{-2}$
as an upper limit to the excess absorption along the line of sight.}} 
We conclude that the ionized gas nebula
does not posses a significant neutral gas component,  
and that there cannot be an extended giant HI envelope along the line of sight. 
 
It is tempting to speculate that the quasar's X-ray emission
is completely absorbed in the {\sl ROSAT} band and we only see
an extended X-ray emission component in the foreground, maybe
from a giant X-ray halo. However,
the X-ray variability of the quasar detected during the present 
and earlier observations (e.g., Pan et al. 1990)
excludes this possibility.

\subsection{High $L_{\rm x}$/$L_{\rm opt}$ and presence of a cooling flow ?}

Given the high ratio of $L_{\rm x}$/$L_{\rm opt}$ of 
MR\,2251-178 (Ricker et al. 1978), is it possible 
that a cooling flow contributes to
its X-ray emission ?
In this context, it is also interesting to note that recently, evidence has been
reported for the presence of highly ionized Oxygen
absorption in cluster cooling flows (e.g., Buote 2000).

We consider the contribution of a cooling flow very unlikely, though, 
because (i) MR\,2251-178 is known 
to be highly variable, (ii) the quasar appears to be 
located far from the center of the cluster,
in its outskirts, and 
(iii) Macchetto et al. (1990) and Shopbell et al. (1999)
present evidence that
the extended [OIII] envelope is not linked to a cooling flow. 

\subsection{X-ray filament between MR2251-178 and the cD galaxy} 

Whereas the bulk of the X-ray emission from MR2251-178
is consistent with a point-source origin, there is some
extended emission at weak emission levels. 
The most interesting feature is a `bridge' between the quasar
and the cD galaxy of the cluster. 

Presently, it is unclear, whether this emission region 
is indeed a filament of extended emission that connects 
the cD galaxy and MR2251-178, or whether it corresponds to several
unresolved point sources, or whether the two sources X23 and X24 (Fig. 1, 2)
are extended.
If real, a giant outflow cone, or the remnant
of a past interaction with the cD galaxy are possible
explanations of the filament.   
Given the present uncertainties, we do not speculate further about 
its origin. 
The feature is potentially
very interesting, though, and should be searched for
with {\sl XMM} and {\sl Chandra} observations.

\subsection{Eastern source excess}

An excess of X-ray sources is detected east of MR2251-178.
We find 19 sources in a region of 20\arcmin$\times$20\arcmin~ 
above a flux level of 2\,10$^{-14}$ erg\,cm$^{-2}$\,s, 
whereas only 4.7 sources are expected according to the 
the $\log N - \log S$ distribution of Hasinger et al. (1994). 
This corresponds to a source overdensity of a factor 4.  
An X-ray source excess, albeit more symmetrically distributed,
was previously noted in, e.g., the NGC\,507 and
NGC\,383 group of galaxies (Kim \& Fabbiano 1995; Komossa \& B\"ohringer 1999, Arp 2001),
and two distant galaxy clusters observed by {\sl Chandra} 
(Elvis et al. 2000, Cappi et al. 2000). Kim \& Fabbiano speculated
that these sources represent cooling clumps in the halo of NGC\,507.
This explanation is unlikely here, because the sources are located far in the outskirts 
of the cluster, and the individual clumps would be exceptionally luminous
(X18: $L_{\rm x} \simeq 9\,10^{42}$ erg/s). 
Optical follow-up observations are required for identification of the X-ray sources
and would then give clues to their origin.

\subsection{Weakness of intra-cluster gas emission}  

Phillips et al. (1980) found MR\,2251-178 to be located in 
the outskirts of a cluster of galaxies. They
counted $\sim$50 member galaxies and provided spectroscopy
of the four brightest. The cluster center is
located close to the brightest elliptical 
(see Fig. 3) which is occasionally referred to as cD galaxy. 

In X-rays, there is a striking absence of bright intracluster gas emission
in contrast to what is usually seen in nearby clusters like this one
(e.g., B\"ohringer et al. 2000).  
We detect only few excess photons in the direction of the optical cluster
center. If all these photons are attributed to ICM emission 
(instead of currently unresolved point sources), we 
derive an X-ray luminosity of $L_{\rm x} \simeq 1.6\,10^{42}$ erg/s.   

The weakness of the ICM X-ray emission suggests, 
that what appears as one single cluster may just
be a chance projection of several smaller clusters or groups. 
In fact, Bergeron et al. (1983) find evidence for a second cluster
at redshift $z=0.12$, based on spectroscopy of one galaxy in the field
of the MR2251-178 cluster. 

The present X-ray results underscore the importance to use X-rays
to distinguish between real clusters and chance projections
(e.g., B\"ohringer et al. 2000), as long
as redshifts for a large number of galaxies in the field are not
available.

\section{Summary and  conclusions}

We have presented an analysis of the soft X-ray properties
of MR\,2251-178 and its host cluster, using \ros PSPC observations. 
The derived mean (0.1--2.4)\,keV  X-ray luminosity of 10$^{45}$ erg/s  
places the quasar among the most X-ray-luminous AGN 
in the local universe. Remarkable for such a luminous source,
we find evidence for an X-ray flaring event with a rise time
of 10\,ksec and an amplitude of a factor $\sim$2.
The combination of earlier optical observations with the present
X-ray data gives the following results: 

An HI extension of the giant HII gas nebula around MR\,2251-178 leads to
the expectation of excess cold absorption in the soft X-ray band which
is, however, not detected ($N_{\rm {H, excess}} < 5\,10^{19}$ cm$^{-2}$).
On the other hand, an {\em ionized} absorber of high column density
is present which, however, is too highly ionized to account for
the optical extended [OIII] emission. 
The shape of the observed X-ray spectrum and the
X-ray variability of the quasar ensure us that we indeed see
X-ray emission from the quasar itself, and not from a giant extended
X-ray halo that may be located outside the X-ray cold absorber. 
At weak emission levels, a filament extends between MR2251-178 and 
the cD galaxy of its host cluster. 
 
East to the quasar there is a significant (factor $\sim$4) excess of X-ray sources 
above the expected  number of background sources. 
The X-ray emission expected from the cluster to which
MR2251-178 belongs is found to be weak or absent. 
 
Deep high-resolution {\sl Chandra} observations have the potential to
solve the enigmas this exceptional source presents to us.  

\begin{acknowledgements}
We thank Gary Ferland for providing {\em{Cloudy}}, 
Wolfgang Brinkmann and Yasushi Ikebe 
for a critical reading of the manuscript, and the referee,
James Reeves, for useful comments. 
The {\sl ROSAT} project was supported by the German Bundes\-mini\-ste\-rium
f\"ur Bildung, Wissenschaft, Forschung und
Technologie 
(BMBF/DLR) and the Max-Planck-Society.
The optical image shown is based on photographic
data obtained using the UK Schmidt Telescope.
The UK Schmidt Telescope was operated by the Royal Observatory
Edinburgh, with funding from the UK Science and Engineering Research
Council, until 1988 June, and thereafter by the Anglo-Australian
Observatory.

Preprints of this
and related papers can be retrieved at \\
http://www.xray.mpe.mpg.de/$\sim$skomossa/

\end{acknowledgements}

\end{document}